\documentclass[aps,prl,twocolumn,showpacs,superscriptaddress,nofootinbib,floats]{revtex4}

\usepackage{epsfig}

\def\be{\begin{equation}}
\def\ee{\end{equation}}

\def\pb{$^{208}$Pb}
\def\hf{$^{170}$Hf}

\begin{document}

\title{Systematics of quadrupolar correlation energies}

\author{M. Bender}
\affiliation{Department of Physics and Institute for Nuclear Theory
Box 351560, University of Washington, Seattle, WA 98195}

\author{G. F. Bertsch}
\affiliation{Department of Physics and Institute for Nuclear Theory
Box 351560, University of Washington, Seattle, WA 98195}

\author{P.-H. Heenen}
\affiliation{Service de Physique Nucl{\'e}aire Th{\'e}orique,
             Universit{\'e} Libre de Bruxelles,
             CP 229, B--1050 Brussels, Belgium}

\date{Oct. 5, 2004}


\begin{abstract}

We calculate correlation energies associated with the 
quadrupolar shape degrees of freedom with a view to improving 
the self-consistent mean-field theory of nuclear binding energies.
The Generator Coordinate Method is employed using mean-field
wave functions and the  Skyrme SLy4 interaction.  Systematic results
are presented for 605 even-even nuclei of known binding energies,
going from mass A=16 up to the heaviest known.  The correlation
energies range from 0.5 to 6.0 MeV in magnitude and are rather
smooth except for large variations at magic numbers and in light
nuclei.  Inclusion of these correlation energies in the calculated
binding energy is found to improve two deficiencies of the Skyrme
mean field theory.  The pure mean field theory has an exaggerated 
shell effect at neutron magic numbers and addition of the correlation
energies reduce it.  The correlations also explain the phenomenon
of mutually enhanced magicity, an interaction between neutron and
proton shell effects that is not explicable in mean field theory.

\end{abstract}
\maketitle

More than 3000 atomic nuclei with different combinations of proton
number $Z$ and neutron number $N$ have been experimentally
identified so far, and for about 2000 of them their masses, or,
equivalently, their binding energies, have been measured
\cite{au03}. This number is only a fraction (about 25\%) of all
the nuclei which are predicted to be bound, i.e. stable against
nucleon emission. Since a large fraction of the unknown nuclei
seems out of reach to be studied experimentally, an accurate theory
of binding energies is needed.  For example, one of the important
motivations for studying nuclei far from stability is to better
understand nucleosynthesis and the astrophysical environment in
which it takes place \cite{rproc}. In particular, the r-process of
heavy-element production leaves an imprint of the binding energies
of far-unstable nuclei on the mass abundances present today.

Among the theories of nuclear binding energy, the self-consistent
mean-field theory~\cite{BHR03} stands out as a fundamentally
justified approach that is also computationally tractable over the
entire mass table. Global calculations of binding energies are now
available including both pairing and deformation effects in the
mean field, using energy functionals of the nonrelativistic Skyrme
form \cite{st03} or a relativistic mean-field model \cite{la99}. A
systematic problem with such theories is an exaggeration of the
increased binding at magic numbers, the shell effect.  The
accuracy of these pure mean field theories is of 1.5-2.0 MeV rms
errors in binding energies \cite{gb04}, much poorer than that of more
phenomenological approaches such as the Finite Range Liquid
Droplet Model \cite{MN95}.

An obvious route to improve the theory is to include correlation
effects.  Indeed, this has already been done in a phenomenological
way with good results \cite{to00,go03}. Some correlation effects
can be subsumed in the parametrization of the mean-field energy
functional, but not all.  Most obviously, the rotational
correlation energy of deformed nuclei is beyond mean-field theory,
which can only describe the intrinsic state of a rotational band.
The requirements on a theory of correlation energies are severe.
Besides being computationally tractable, it must be systematic,
applicable to both spherical and deformed nuclei.  One approach,
the RPA theory of correlation energies, has been applied recently
to the binding energy problem by Baroni {\it et al.}, \cite{BA04}.
However, the RPA is not well convergent for short-range
interactions, limiting its usefulness \cite{SJ02}.

In this letter we report on global calculations of correlation
energies in another well-defined theoretical extension of the
mean-field approximation~\cite{BHV03}, based on the Generator
Coordinate Method (GCM). The quadrupolar fluctuations about the
mean-field solution are determined variationally by mixing
configurations around the mean-field ground state. Our method
includes also a projection on good angular momentum and particle
number and applies to all nuclei irrespective of the shape
characteristics of their ground states. It is thus systematic,
which is an important criterion for constructing mass tables. It
also satisfies the goal of including in the correlation energy,
the rotational energy of deformed states.

The technical details of our calculation are as follows.  The
self-consistent mean-field equations are solved using the method
presented in ref. \cite{BF85}.  Here the wave functions are
represented on a triaxial 3-dimensional spatial grid, rather than
in an oscillator basis as was done in ref. \cite{st03,la99}.  The effective
nucleon-nucleon interaction is an energy density functional of the
Skyrme form; we have used the SLy4 parameter set \cite{CB98}. This
set was fitted to nuclear matter properties, nuclear charge radii,
as well as binding energies of doubly magic nuclei. We also
included a local pairing interaction of the form used in ref.
\cite{RB99}, but with a reduced strength of $v_0 = -1000$
MeV fm$^3$ to compensate the explicit correlations added by the
generator coordinate treatment\cite{BHV03}.

The GCM has as ingredients a set of wave functions generated by
constrained mean-field calculations and a Hamiltonian for
calculating matrix elements among those configurations. The
constraint that we use is the quadrupole operator $Q(r)$ and we
minimize the function:
\be E[\phi] - \lambda  \int d^3 r
\, \rho(r) Q(r).
\ee
Here $\rho$ is the mass density and $\lambda$ is a Lagrange
multiplier adjusted to get a specific value of the quadrupole
moment $q=\int d^3 r \rho(r) Q(r)$. The value of $q$ is positive
for prolate and negative for oblate deformations.  The choice of
the configurations $|q\rangle$ is a computational issue that has
to be carefully addressed for systematic calculations.  If states
are too closely spaced there is a large redundancy as well as
an excessive computational cost.
We have tailored our selection of configurations for a target
accuracy of 0.2 MeV in the correlation energies. In ref.
\cite{bo90}, it was found that this accuracy can be achieved when
overlap probabilities of neighboring configurations is about 0.5
or higher, and we adopt that criterion here.  We also restrict the
deformations to shapes that have energies within 5 MeV of the
mean-field ground state. Typically this gives $n_q=7-20$ different
configurations in our computational space \cite{be04}.

There is an ambiguity in the choice of the Hamiltonian $\hat H$  for use with
a mean-field energy functional \cite{du03}.  We use the so-called mixed
density for the calculation of the density-dependent term in the
Skyrme Hamiltonian, as is mostly done in the literature \cite{fn1}.
The GCM energy is the
minimum of the expression
$$
\langle \Psi| \hat H |\Psi \rangle \over \langle\Psi | \Psi\rangle
$$
where $|\Psi\rangle$ is an arbitrary state in the basis of the
$|q\rangle$. The minimization is performed by solving the
corresponding discretized Hill-Wheeler equations \cite{hi53}. The
orientation of the intrinsic deformed states $|q\rangle$ could be
considered a continuous degree of freedom to be included in the
minimization, but in practice we achieve the same result by
projecting the intrinsic states on angular momentum zero when
calculating matrix elements.  The theoretical correlation energy
is then obtained in two steps: we determine first the energy gain
of an angular-momentum projected configuration $|q\rangle$ with
respect to the mean-field minimum, $E_{J=0}$, and after that, the
additional energy gained by mixing configurations with different
$q$ magnitudes, $E_{HW}$.

The needed matrix elements $\langle q  | q' \rangle$ and $\langle
q | \hat H | q' \rangle $ are calculated with the computer code
described in ref. \cite{BHV03}.  Besides performing the rotations
to calculate the matrix elements needed for angular momentum
projection, the code carries out a projection of the BCS wave
function on good particle number $N$ and $Z$.  The number
projection and the rotation operation are computationally very
demanding, and it was necessary to find additional computational
shortcuts to make the global study feasible. One important savings
was the use of a topological Gaussian overlap approximation which
allows a two-point evaluation of the angular momentum projection
integrals \cite{ha02,be04}. The top-GOA was found to be quite
adequate, except for very light nuclei,
where a three-point approximation was sometimes needed. Another
place where we made considerable computational savings was in the
calculation of off-diagonal matrix elements in the $|q\rangle$
space. In principle, an $n$-dimensional basis requires the
calculation of $n(n+1)/2$ overlaps and Hamiltonian matrix elements. These
must be done explicitly for neighboring configurations, but matrix
elements for more distant configurations can be estimated using
another Gaussian Ansatz based on a measure of the separation
between the configurations\cite{bo90}. A tricky point arises in the mixing of
weak prolate and oblate deformations which can have very high
overlaps. We deal with this problem by defining a two-dimensional
metric for calculating the separation of configurations as
described in ref. \cite{be04}. With these approximations, the computational effort
for the GCM minimization was reduced by 1 to 2 orders of
magnitude, but still requiring on the order of $10^{17}$ floating
point operations for the entire table of 605 nuclei.

\begin{figure}[bht]
\centerline{\epsfig{figure=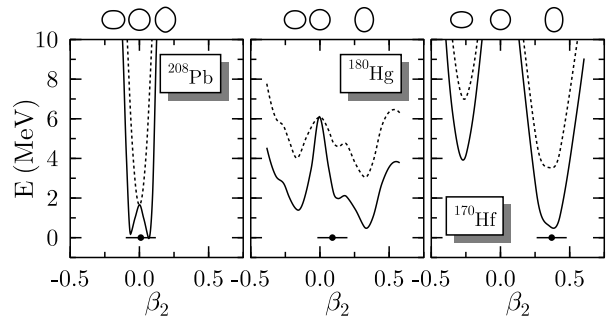,clip=}}
\caption{\label{fig1} Typical potential energy landscapes in the
deformation coordinate.  Shown are the curves for 
\pb, $^{180}$Hg, \hf, representing a magic nucleus, a soft nucleus,
and a deformed nucleus, respectively.  See text for explanation.
}
\end{figure}

We now present three calculations, typical of three different
topologies. In Fig.\ref{fig1} are plotted the deformation energy
curves of a doubly magic nucleus, $^{208}$Pb, a statically
deformed nucleus $^{170}$Hf, and a nucleus with oblate and prolate
minima close in energy, $^{180}$Hg. The configuration spaces used
are given in Table 1. The $^{208}$Pb is a very stiff nucleus, and
only needed 4 configurations to achieve the target accuracy. In
contrast, $^{180}$Hg is a rather soft nucleus, and it required 15
configurations ranging from -20 b to +32 b to map out the
accessible deformations. The mean-field potential energy landscapes for these
three nuclei are shown as the dashed lines in Fig.~\ref{fig1}.
\begin{table}
\caption{Configuration spaces for typical heavy
nuclei.  Mass quadrupole moments $q$ are given in units of barns.
Oblate and prolate configurations are listed on separate lines.
}
\begin{tabular}{c|c|c}
\hline
Nucleus   &  $n_q$& $q$ values  \\
\hline
$^{208}$Pb & 4& -10 -5\\
& & +5 +10\\
\hline
$^{180}$Hg &15& -20 -16 -14 -10 -6 -4\\
&& +4 +6 +8 +12 +16 +20 +24 +28 +32\\
\hline
$^{170}$Hf &12&-20 -16 -13.75 -10 -5\\
&& 5 +10 +15 +19.25 +22 +25 +30\\
\end{tabular}
\end{table}
\begin{table}
\caption{Quadrupolar deformation and correlation energies of typical heavy
nuclei.  Energies are given in MeV.
}
\begin{tabular}{c|cccc}
\hline
Nucleus   &$ E_{def}$ & $ E_{J=0}$ &
$ E_{HW}$   & $ E_{corr} $\\
\hline
$^{208}$Pb & 0.0 &-1.7  &0.0  &-1.7  \\
$^{180}$Hg &-3.0 &-2.5 &-0.5 & -3.1 \\
$^{170}$Hf & -12.2 &-2.9 & -0.4 & -3.4 \\
\end{tabular}
\end{table}
One sees the narrow minimum at zero deformation for the magic
$^{208}$Pb, a much flatter potential energy surface for
$^{180}$Hg, and a surface with a strongly deformed minimum for
$^{170}$Hf.  The energy gain from angular momentum projection  is
shown by the solid lines in the table.  Obviously, there is no
change  for the spherical configurations but for finite
$|q\rangle$ the projection lowers the energy below the mean-field
value for every $q$.  Even for $^{208}$Pb, there is a net energy
gain $E_{J=0}$ of the lowest projected energy with respect to the
mean-field ground state.  These numbers are given in Table 2.

The second step in our calculation is to apply the GCM to mix
configurations with different values of $q$. The additional energy
$E_{HW}$ is given in Table 2 and shown as a dot in Fig. 1. The
horizontal position of the dot indicates the average deformation
of the mixed wave function (see~\cite{BHV03} for its precise
definition).
\begin{figure}[bht]
\centerline{\epsfig{figure=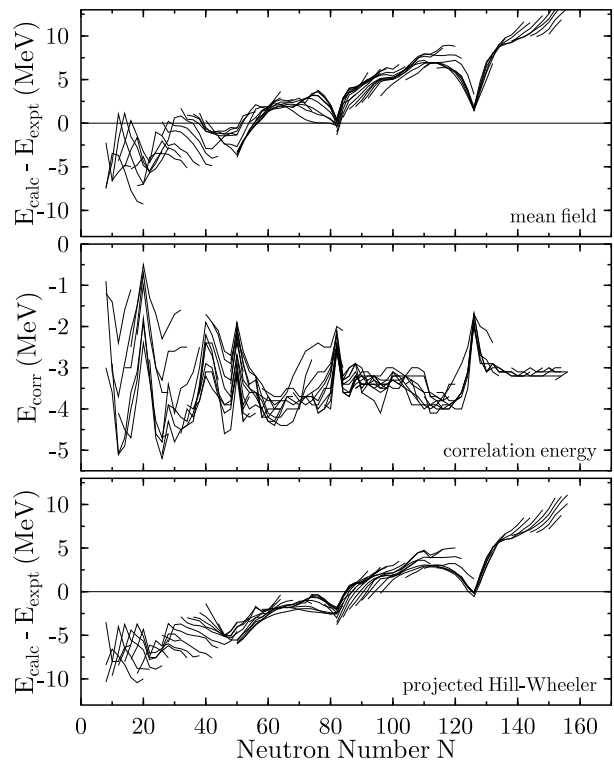}} \caption{\label{fig4}
 Top panel: Difference of experimental and mean-field binding energies for
the SLy4 interaction \protect\cite{st03}, as a function of neutron number
with lines connecting isotopic chains. Middle panel: Theoretical quadrupolar
correlation energies of the even-even nuclei. Note the expanded energy
scale. Bottom panel: Difference of experimental and theoretical binding
energies including the correlation energy. Comparison between binding
energies obtained with and without the correlation contribution. }
\end{figure}

Our calculations include all but the lightest even-even nuclei for
which the mass is known.  The results presented in Fig. 2, are
displayed as a function of neutron number $N$ with isotopes
connected by lines.  On the top panel, we show the energy
difference between the SLy4 mean-field theory and experiment\cite{fn1}. Here
the shell effect is very prominent at neutron numbers $N=50$, 82,
and 126.  The middle panel shows the calculated quadrupolar
correlations energies. One sees that, for most nuclei, it is
between 3 and 4 MeV, irrespective of whether the nucleus is
spherical or deformed.  However, near doubly magic nuclei, the
correlation energy is much smaller.  This difference has the right
sign to mitigate the too-strong shell effect of the mean-field
theory. The bottom panel in Fig. 2 shows the binding energies with
and without inclusion of the correlation energy.  There is a
distinct improvement, but one sees that there remains some
residual shell effect. Since we have only included the axial
quadrupolar field, there could be significant correlation energies
arising from other kinds of deformation or from pairing vibrations
around closed shells\cite{BA04}. 
Also, it is clear that a more complete theory requires a refitting
of the energy functional to take into account the contribution
from the correlation energy.

\begin{figure}[bht]
\centerline{\epsfig{figure=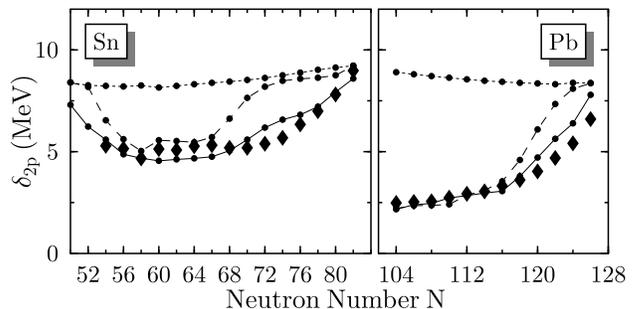}}
\caption{\label{fig5} Two-proton gaps for Pb and Sn isotopic chains.}
\end{figure}

A feature of the binding systematics that has been difficult to
understand in mean-field theory is ``mutually
enhanced magicity" \cite{LMP03}, the enhancement of shell effects when
both neutrons and protons form closed shells.  This behavior
can be seen in the neutron dependences of the two-proton gap,
defined as
\begin{equation}
\label{eq:d2p}
\delta_{2p}(N,Z)
= E (N,Z-2) - 2 E(N,Z) + E(N,Z+2)
.
\end{equation}
In the independent-particle shell model, $\delta_{2p}$ is
proportional to the difference between the Fermi energies of
nuclei differing by two protons and thus it is a measure of shell
gaps. This quantity is plotted in Fig. 3 for the $Z=50$ (Sn) and
$Z=82$ (Pb) proton gaps, with the experimental points shown by black
diamonds. There is clearly a reduction of $\delta_{2p}$ when going
away from the doubly magic $^{132}$Sn with 82 neutrons and doubly
magic $^{208}$Pb with 126 neutrons. Neglecting deformations, the
mean-field prediction for $\delta_{2p}$ is quite flat, as shown by
the short-dashed lines on the graphs.  This reflects the
independence of the single-particle proton gap on the neutrons in that
approximation\cite{MB02}. Allowing deformations in the mean-field energies
already gives a change into the right direction, as shown by the
long-dashed lines. The calculation including the full quadrupole
correlation energy is shown by the solid lines.  One sees that it
brings the calculations even closer to experiment. The correlation
energy thus provides a plausible explanation of mutually enhanced
magicity. This conclusion is corroborated by a recent calculation
of $\delta_{2p}$ in the Sn chain using the Bohr Hamiltonian,
showing also the large improvement brought by quadrupolar
correlations \cite{fl04}.

In conclusion, we find that quadrupole correlations beyond the
mean field capture a significant part of the binding energy that
is not describable in mean-field theory. We have shown that global
calculations of these correlations are today feasible and could be
incorporated in the future in the construction of effective
interactions. Further improvements should be obtained by including
other correlation effects related to octupole deformations or
pairing vibration~\cite{BA04,he02}. Our basis does not contain triaxial
shapes, but we do not believe they play as important a role. One
of the major challenges which remains open is to include
neutron-proton pairing effects, which may be responsible for the
scatter of the residuals around the light $N=Z$ nuclei in Fig. 2.

This work is supported by the U.S. Dept. of Energy under Grant
DE-FG02-00ER41132 and the Belgian Science Policy Office under
contract PAI P5-07. The computations were performed at the
National Energy Research Scientific Computing Center, supported by
Department of Energy under Contract No. DE-AC03-76SF00098.
PHH thanks the Institute for Nuclear Theory for hospitality where
part of this work was carried out, and we also thank B. Sabbey for
help.


\begin{thebibliography}{99}

\bibitem{au03}
G. Audi, A. H. Wapstra, and C. Thibault,
Nucl. Phys. \textbf{A729}, 337 (2003);
the data file is available at
{\tt http://www-csnsm.in2p3.fr/AMDC/masstables/\\
Ame2003/mass.mas03}.

\bibitem{rproc} K. Langanke and M. Wiescher,
  Rep. Prog. Phys. \textbf{64}, 1657 (2001).

\bibitem{BHR03}
  M. Bender, P.-H. Heenen, and P.-G. Reinhard,
  Rev. Mod. Phys. \textbf{75}, 121 (2003).


\bibitem{st03}
  M. V. Stoitsov, J. Dobaczewski, W. Nazarewicz, S. Pittel, and D. J. Dean,
  Phys. Rev. C \textbf{68}, 054312 (2003);
  {\tt http://www.fuw.edu.pl/\~{}dobaczew/\\
thodri/s9l20v56.rev.txt}

\bibitem{la99} G.A.~Lalazissis, S.~Raman, and P.~Ring, At. Data Nucl.
Data Tables  {\bf 71 } 1 (1999).

\bibitem{gb04} G.F.~Bertsch,B.~Sabbey, and M.~Uusn\"akki, to be published.

\bibitem{MN95}
  P. M{\"o}ller, J. R. Nix, W. D. Myers, and W. J. Swiatecki,
  Atom. Data Nucl. Data Tables \textbf{59}, 185 (1995).

\bibitem{to00}
F. Tondeur,  S. Goriely, J. M. Pearson, and M. Onsi,
Phys. Rev. \textbf{62} 024308 (2000).

\bibitem{go03}
S. Goriely, M. Samyn, M. Bender, and J. M. Pearson,
Phys. Rev. C \textbf{68}, 054325 (2003).

\bibitem{BA04}
S. Baroni, M. Armati, F. Barranco, R. A. Broglia, G. Colo', G. Gori,
and E. Vigezzi,
arXiv:nucl-th/0404019

\bibitem{SJ02}
I. Stetcu and C. Johnson,
Phys. Rev. C \textbf{66}, 034301 (2002).

\bibitem{BHV03}
M. Bender and P.-H. Heenen, Nucl. Phys. \textbf{A713} (2003) 390;
A. Valor, P.-H. Heenen and P. Bonche, Nucl. Phys. \textbf{A671}
(2000) 145.

\bibitem{BF85}
P. Bonche, H. Flocard, P.-H. Heenen, S. J. Krieger and M. S. Weiss,
Nucl. Phys. \textbf{A443} (1985) 39.

\bibitem{CB98}
E. Chabanat, P. Bonche, P. Haensel, J. Meyer, and R. Schaeffer,
Nucl. Phys. \textbf{A635}, 231    (1998),
Nucl. Phys. \textbf{A643}, 441(E) (1998).

\bibitem{RB99}
C. Rigollet, P. Bonche, H. Flocard, and P.-H. Heenen,
Phys. Rev. C \textbf{59}, 3120 (1999).

\bibitem{bo90} P. Bonche, et al., Nucl. Phys. {\bf A510} 466 (1990).

\bibitem{be04}
M. Bender, G. F. Bertsch and P.-H. Heenen,
Phys. Rev. C \textbf{69} (2004) 034340.

\bibitem{du03}
T. Duguet and P. Bonche,
Phys. Rev. C \textbf{67} 054308 (2003).


\bibitem{fn1}  We also omit some small terms that appear in the
Hamiltonian formulation of the Skyrme interaction but not in the
energy density functional.

\bibitem{hi53}
D. L. Hill and J. A. Wheeler,
Phys. Rev. \textbf{89} (1953) 1102;
J. J. Griffin and J. A. Wheeler,
Phys. Rev. \textbf{108}, 311 (1957).



\bibitem{ha02}
K. Hagino, P.-G. Reinhard, and G. F. Bertsch,
Phys. Rev. C \textbf{65}, 064320 (2002).

\bibitem{fn2} Since the SLy4 interaction was fitted to only magic nuclei,
the other nuclei show a trend of increasing residuals with mass number.
This is easily removed by a minor adjustment of parameters.

\bibitem{LMP03}
D. Lunney, J. M. Pearson, and C. Thibault,
Rev. Mod. Phys. \textbf{75}, 1021 (2003).

\bibitem{fl04} P. Fleischer, et al., ArXiv:nucl-th/0408032 (2004).


\bibitem{he02} P.-H.~Heenen, et al., Eur. Phys. J. A {\bf 11 } 393 (2001).

\bibitem{MB02} M. Bender, et al., Eur. Phys. J. A {\bf 14} 23 (2002).
\end{thebibliography}
\end{document}